\documentclass[epj]{svjour}
\usepackage{graphics}
\usepackage[]{youngtab}
\begin{document}
\title{Excited Baryons in Large $N_c$ QCD}
\subtitle{Matching the $1/N_c$ expansion to quark models using the permutation group $S_N$\\}
\author{D.~Pirjol\inst{1} \and C.~Schat\inst{2}
}                     
%
%
\institute{National Institute for Physics and Nuclear Engineering, 
Department of Particle Physics, \\ 077125 Bucharest, Romania. \and CONICET and Departamento de F\'{\i}sica, FCEyN, Universidad de Buenos Aires,
\\ Ciudad Universitaria, Pab.1, (1428) Buenos Aires, Argentina.}
\date{Received: date / Revised version: date}
%
\abstract{ 
We show how to match quark models to the $1/N_c$ expansion of QCD. As an example we discuss in detail the mass operator 
of orbitally excited baryons and match it to the one-gluon exchange and the 
one-boson exchange variants of the 
quark model.  
The matching procedure is very general and makes use of the transformation properties
of states and operators under $S_N^{\rm orb} \times S_N^{\rm sp-fl}$, the
permutation group acting on the orbital and  spin-flavor degrees of freedom
of $N$ quarks.
} 
\maketitle
\section{Introduction}
\label{intro}
Quark models provide a simple and  intuitive picture for the physics of 
ground state baryons and their excitations. These models constitute 
 important 
assets in the toolbox of the hadronic physicist. It is of great relevance to 
sharpen up tools that have proven to be useful and put them 
on a sounder ground, establishing the connection with the underlying theory 
of the strong interactions. Hopefully this will shed light on why these simple models work, something that 
can also hint at how to improve them.  

The large $N_c$ limit of QCD provides a leading order picture for hadrons which can be corrected systematically including $1/N_c$ corrections  \cite{Dashen:1993jt}. This program 
can be realized in terms of a quark operator expansion, which gives rise 
to a physical picture similar to the one of the phenomenological quark models, but is closer connected 
to QCD.  In this context quark models gain additional significance. 
Here we want to show how it is possible to match these quark models 
to the more general  $1/N_c$ expansion of QCD. It turns out that classifying the transformation properties of states and operators under permutations using the symmetric group $S_N$ is the key to perform this matching explicitly \cite{Pirjol:2007ed}. For $N_c=3$ the symmetric group $S_3$ has been used previously in Ref.~\cite{Collins:1998ny} to compare the predictions of two 
versions of the quark model, one based on   
one-gluon exchange \cite{Isgur:1978xj} and one based on one-boson exchange 
 \cite{Glozman:1995fu}. Here we generalize this to arbitrary $N_c$ 
and compare both models with the more general $1/N_c$ expansion.

Before getting into the details of the matching procedure it is useful  to
recall a few general points that make the large number of colors limit
interesting and useful:
\begin{itemize}
\item Although in the large $N_c$ limit the number of
degrees of freedom increases, the physics simplifies.

\item The $1/N_c$ expansion is the only candidate for a perturbative expansion of QCD at all energies.

\item In the $N_c \rightarrow \infty$ limit baryons fall into
 irreducible representations of the {\it contracted} spin-flavor algebra $SU(2 n_f)_c$ ,
also known as ${\cal K}$-symmetry, that relates properties of states in different
multiplets of flavor symmetry.

\item The breaking of spin-flavor symmetry can be studied order by
order in $1/N_c$ as an operator expansion.

\end{itemize}

The successful applications of the $1/N_c$ expansion to the study of
ground state baryons make the excited baryons  especially 
interesting because they provide a wider testing ground for the $1/N_c$
expansion. 
It is important to 
stress that already at leading order in the large $N_c$ limit it is possible to obtain 
significant qualitative insights into the structure of excited baryons, among which we would like to highlight the 
following:

\begin{itemize}

\item  The three towers \cite{Pirjol:1997bp} \cite{Pirjol:2003ye} \cite{Cohen:2003tb} predicted by ${\cal K}$-symmetry 
for the $L=1$ negative parity $N^*$ baryons, labeled by ${\cal K}=0,1,2$ with ${\cal K}$
related to the isospin $I$ and spin $J$ of the $N^*$'s by $I+J \ge {\cal K}\ge|I-J|$. 

\item The vanishing of the strong decay width  $\Gamma(N^*_{\frac12} \to [N \pi]_S)$
for $N^*_{\frac12}$ in the ${\cal K}=0$ tower, which provides a 
natural explanation for the relative suppresion of pion decays for the  
$N^*(1535)$  \cite{Pirjol:1997bp} \cite{Cohen:2003tb} \cite{Pirjol:2003qk}.

\item The order ${\cal O}(N_c^0)$ mass splitting of the $SU(3)$ singlets   $\Lambda(1405)$ - $\Lambda(1520)$
in the $[\mathbf{70},1^-]$ multiplet \cite{Schat:2001xr}. 

\end{itemize}

The $1/N_c$ expansion has also been applied to orbitally excited baryons
up to subleading order in $1/N_c$ 
\cite{Goity:1996hk,Pirjol:1997bp,Carlson:1998vx,Pirjol:2003ye,Schat:2001xr,GSS03,Pirjol:2006ne,Matagne:2004pm}.
The construction of the states and operators for excited states presented in 
Ref.~\cite{Carlson:1998vx} has been inspired by the quark model
picture of the excited states and makes use of the
decomposition of the spin-flavor states into ``core'' and ``excited'' quark subsystems. A formal justification of this in terms 
of a symmetry argument can be obtained considering the permutation group $S_N$ \cite{Pirjol:2007ed}. In several recent papers \cite{Matagne:2006dj,Semay:2007cv} the validity of 
the usual approach based on the core+excited quark decomposition has been questioned, 
and this symmetry argument settles the objections raised in these works.

Including also the orbital degrees of freedom, 
the complete permutation symmetry is $S_N \subset S_N^{orb} \times S_N^{sp-fl}$,
the diagonal subgroup of the permutations acting on both orbital and spin-flavor
degrees of freedom. Of course, although $S_N$ is a good symmetry of the 
quark model Hamiltonian, $S_N^{sp-fl}$ is not, and mixing between different
irreps can occur in general (configuration mixing).

\begin{table*}[t]
\caption{Irreps of the permutation group $S_N$ used in the text. The Young diagrams
shown correspond to $N=5$.}
\label{tab:1}  
\begin{center}
\begin{tabular}{ccccc}
\hline \noalign{\smallskip}
irrep : & $S$ & $MS$ & $E$ & $A'$ \\
\hline \noalign{\smallskip}
 & \,\raisebox{-0.1cm}{\yng(5)}\, & \,\raisebox{-0.3cm}{\yng(4,1)}\, & 
\,\raisebox{-0.3cm}{\yng(3,2)}\, & \,\raisebox{-0.6cm}{\yng(3,1,1)}\, \\
\noalign{\smallskip}
partition:  & $[N]$      & $[N-1,1]$ & $[N-2,2]$ & $[N-2,1,1] $ \\
\noalign{\smallskip}
dim  : & 1 &  $N-1$  &  $N(N-3)/2$  &  $(N-1)(N-2)/2 $ \\
\noalign{\smallskip}
\hline
\end{tabular}
\end{center} 
\end{table*}

We consider here in some detail the states transforming in the 
$[N-1,1]$ ($MS$) irrep of $S_N^{\rm sp-fl}$.  
The matrix elements $\langle MS |  \hat O | MS \rangle$ of any operator 
$\hat O$ on mixed symmetric spin-flavor states $MS$ can transform as:
\begin{eqnarray}\label{MS^2}
MS \otimes MS &=& S \oplus MS \oplus A' \oplus E  \ . 
\end{eqnarray}

These irreps of $S_N$ are identified by a partition $[\{n_i\}]$
corresponding to a Young diagram with $n_1. n_2, \cdots$ boxes in each row, see 
Table \ref{tab:1} for notation and typical  Young diagrams for $N=5$.

The construction of the states with correct permutation symmetry under $S_N$
is described in Sec.~\ref{Sec:MSstates}. 
As an example of the decomposition Eq.~(\ref{MS^2})
we show in detail in Sec.~\ref{Sec:4}
the explicit decomposition of the  spin-spin two-body
quark-quark interaction for the one-gluon exchange model, into irreps of 
$S_N^{orb} \times S_N^{sp-fl}$. Finally, the derivation of the mass operator
and the matching onto the $1/N_c$ expansion are presented in Sec.~\ref{Sec:5}.
A complete discussion can be found in Ref.~\cite{Pirjol:2007ed}.

\section{The $MS$ states and their relation to CCGL}
\label{Sec:MSstates}

A basis for the MS spin-flavor wave function can be constructed using the method of
the Young operators (for further details and examples for $N=5$ see 
Ref.~\cite{Pirjol:2007ed}). It can be chosen as the set of $N-1$ 
wave functions, with $k=2, 3, \cdots, N$
\begin{eqnarray}\label{phidef}
\phi_k &=& |q_k\rangle \otimes |im\alpha\rangle_{N-1} - 
|q_1\rangle \otimes |im\alpha\rangle_{N-1}
\end{eqnarray}
where $|q_k\rangle$ denotes the spin-flavor state of the quark $k$, and 
$|im\alpha\rangle_{N-1}$ denotes the spin-flavor state of the subset of $N-1$ quarks (`core')
obtained by removing
quark $k$ from the $N$ quarks. The latter states are symmetric under any permutation
of the $N-1$ quarks. The states $\phi_i$ are not orthogonal, and have the scalar products 
\begin{eqnarray}\label{phinorm}
\langle \phi_i | \phi_j \rangle = S_{ij}\,, \qquad
S_{ij} = \left\{
\begin{array}{cc}
2 & , \quad i=j \\
1 & , \quad i\neq j \\
\end{array}
\right.
\end{eqnarray}

The basis states $\phi_k$ have the following transformation properties under the
action of the transpositions $P_{ij}$ (exchange of the quarks $i,j$)
\begin{eqnarray}\label{phitransp}
P_{1j} \phi_k &=& 
\left\{ 
\begin{array}{cc}
- \phi_k & , j=k \\
\phi_k - \phi_j & , j\neq k \\
\end{array}
\right. \\ 
P_{ij} \phi_k &=& \phi_k \mbox{ if } (i,j) \neq k,1 \\ \label{phitransp2}
P_{ik} \phi_k &=& P_{ki} \phi_k = \phi_i \quad  \mbox{ if } i\neq1 \,.
\end{eqnarray}
These transformation relations can be obtained defining a basis for
the $MS$ irrep of $S_N$ in a general way that is valid for states and operators,   namely, as $\phi_k = (P_{1k} - 1) \phi$ with any $\phi$ satisfying 
$\phi = P_{ij} \phi$ for $i,j=2...N$.

For the orbital wave functions  we adopt a Hartree representation, in 
terms of one-body wave functions $\varphi_s(r)$ for the ground state orbitals, and
$\varphi_p^m(r)$ for the orbitally excited quark. The index $m=\pm 1, 0$ denotes 
the projection of the orbital angular momentum $\vec L$ along the $z$ axis.
The Young operator basis for the
orbital wave functions is given by (with $k=2,3, \cdots, N$)
\begin{eqnarray} \label{chidef}
\chi_k^m(\vec r_1, \cdots , \vec r_N) = 
(P_{1k} - 1) 
\varphi_p^m(\vec r_1) \Pi_{i\neq 1}^N \varphi_s(\vec r_i) \ ,
\end{eqnarray}
and have the same transformations under transpositions
as the spin-flavor basis functions Eqs.~(\ref{phitransp}-\ref{phitransp2}).

The complete spin-flavor-orbital wave function of a baryon $B$ with 
mixed-symmetric spin-flavor symmetry
is written as the $MS \times MS \to S$ inner product of the two basis wave 
functions, for the orbital and spin-flavor components, respectively
\begin{eqnarray}\label{MS}
|B\rangle = \sum_{k,l=2}^N \phi_k \chi_l^m M_{kl}
\end{eqnarray}
The matrix of coefficients $M_{ij}$ are the Clebsch-Gordan
coefficients for the $MS \times MS \to S$ inner product of two irreps of $S_N$.
They can be determined by requiring that the state Eq.~(\ref{MS}) is left
invariant under the action of transpositions acting simultaneously on the
spin-flavor and orbital components. In the $MS$ basis defined by the
transformations Eqs.~(\ref{phitransp}) the matrix $M_{ij}$ is 
\begin{eqnarray}\label{Mmatrix}
\hat M = \left(
\begin{array}{c
cccc}
1 & -\frac{1}{N-1} & -\frac{1}{N-1} & \cdots & -\frac{1}{N-1} \\
 -\frac{1}{N-1} & 1 &  -\frac{1}{N-1} & \cdots &  -\frac{1}{N-1} \\
 -\frac{1}{N-1} &  -\frac{1}{N-1} & 1 & \cdots &  -\frac{1}{N-1} \\
\cdots & \cdots & \cdots & \cdots & \cdots \\
 -\frac{1}{N-1} &  -\frac{1}{N-1} &  -\frac{1}{N-1} & \cdots & 1 \\
\end{array}
\right)
\end{eqnarray}
Since 
any permutation can be represented as a product of transpositions, 
the state given in Eq.~(\ref{MS}) transforms indeed in the $S$ irrep of the 
overall $S_N$ group.

The state Eq.~(\ref{MS}) can be related to the $MS$ states 
constructed
in Ref.~\cite{Carlson:1998vx}, and commonly used in the literature in the 
context of the $1/N_c$ expansion. These states are constructed as tensor
products of an `excited' quark whose identity is fixed as quark $1$, with a 
symmetric `core' of $N-1$ quarks. We will refer to these states as CCGL states.
With this convention, the wave function used in Ref.~\cite{Carlson:1998vx}
has the form
\begin{eqnarray}
|CCGL \rangle = \Phi(SI) \,\, \varphi_p^m(\vec r_1) \Pi_{i=2}^N 
\varphi_s(\vec r_i)
\end{eqnarray}
where $\Phi(SI)$ denotes the spin-flavor component, and the remainder is the 
orbital
wave function in Hartree form. By construction, the spin-flavor wave function
$\Phi(SI)$ transforms in the $MS$ irrep of $SU(4)$, and thus also like
$MS$ under $S_N$. It is symmetric under any exchange of the core quarks,
$P_{ij} \Phi(SI) = \Phi(SI)$ for $i,j\neq 1$. These two properties identify it uniquely 
in terms of the $MS$
basis functions defined in Eq.~(\ref{phidef}) as
$\Phi(SI) = \frac{1}{\sqrt{N(N-1)}}\sum_{k=2}^N \phi_k$. The normalization factor
is chosen such that the state is normalized as $\langle \Phi(SI)|\Phi(SI)\rangle 
= 1$.
Symmetrizing under the `excited' quark index $i=1,2, \cdots , N$, one finds
for the properly normalized symmetric state
\begin{eqnarray}
&& | CCGL \rangle \to \frac{1}{\sqrt{N}}\Sigma_{i=1}^N P_{1i} |CCGL\rangle   =  \nonumber \\ 
&& \!\!\!\!\!\!\!\!\!\! =
\frac{1}{N \sqrt{N-1}} \Sigma_{i=2}^N \Sigma_{j=2}^N (P_{1i} \phi_j) \chi_i  = 
- \frac{\sqrt{N-1}}{N} |B\rangle , 
\end{eqnarray}
where the terms with different $i$ in the sum are orthogonal states.
This gives the relation between the CCGL state and the 
symmetric state constructed above in Eq.~(\ref{MS}).

\section{OGE and OBE quark model interactions}

In this Section we present two representative quark models that will be 
matched to the $1/N_c$ expansion. 
We start considering
 the one-gluon exchange potential (OGE),
which follows from a perturbative expansion in $\alpha_s(m_Q)$ in the heavy quark limit
$m_Q \gg \Lambda_{QCD}$.

We consider a Hamiltonian containing a spin-flavor symmetric term $H_0$ 
(the confining potential and kinetic terms), plus spin-isospin dependent 
two-body interaction terms $V_{ij}$ 
\begin{eqnarray}\label{Hamilton}
\!\!\!\!\! H = H_0 + g^2 \sum_{i<j} \frac{\lambda_i^a}{2} \frac{\lambda_j^a}{2} V_{ij} \to
H_0 - g^2\frac{N_c+1}{2N_c}\sum_{i<j} V_{ij}
\end{eqnarray}
where $\lambda^a$ are the generators of SU(3) color in the fundamental representation.
The second equality holds on color-singlet hadronic states, on which
the color interaction $t_i^a t_j^a$ evaluates to the color factor 
$(N_c+1)/(2N_c)$.
We will restrict ourselves only to color neutral hadronic states.

In the nonrelativistic limit, the two-body interaction $V_{ij}$ contains three 
terms: the spin-spin interaction $(V_{ss})$, the quadrupole interaction 
$V_q$ and the spin-orbit terms $V_{so}$. 
We write these interaction terms following Ref.~\cite{Collins:1998ny}, where 
$f_{0,1,2}(r_{ij})$ are unspecified functions of the interquark distances
\begin{eqnarray}
\label{vssoge}
V_{ss} &=& \sum_{i < j=1}^N f_0(r_{ij}) \vec s_i \cdot \vec s_j \\
V_{q} &=& \sum_{i < j=1}^N f_2(r_{ij}) \Big[3(\hat r_{ij} \cdot \vec s_i)(\hat r_{ij} \cdot \vec s_j)
 -  (\vec s_i \cdot \vec s_j) \Big] \\ 
 \label{vsooge}
V_{so} &=& \sum_{i < j=1}^N f_1(r_{ij}) \Big[ 
(\vec r_{ij} \times \vec p_i ) \cdot \vec s_i - (\vec r_{ij} \times \vec p_j ) \cdot \vec s_j \\ 
& & \hspace{1.5cm}+ 2 (\vec r_{ij} \times \vec p_i ) \cdot \vec s_j - 2 
(\vec r_{ij} \times \vec p_j ) \cdot \vec s_i \Big] \nonumber
\end{eqnarray}

We consider also the mass operator of the orbitally excited
baryons in a second model for the quark-quark interaction. 
In Ref.~\cite{Glozman:1995fu} it was suggested that
pion-exchange mediated quark-quark interactions can reproduce better
the observed mass spectrum of these states. The physical idea is that 
at the energy scales of quarks inside a hadron, the appropriate degrees
of freedom are quarks, gluons and the Goldstone bosons of the broken chiral
group $SU(2)_L \times SU(2)_R \to SU(2)$ \cite{Manohar:1983md}.
The exchange of Goldstone bosons changes the short distance form of the
quark-quark interactions, and introduces a different spin-flavor structure.
The new potentials $\tilde V_{ij}$ are obtained from Eqs.(\ref{vssoge})-(\ref{vsooge}),   with the replacements
$f_{0,1,2}(r_{ij}) \rightarrow g_{0,1,2}(r_{ij}) t^a_i t^a_j $ 
where the isospin generators are $t^a = \frac12 \tau^a$, and $g_{0,1,2}(r_{ij})$
are unspecified functions.

The interaction Hamiltonian of the one boson exchange (OBE) model has
 then the form
\begin{eqnarray}\label{HamiltonGR}
H = H_0 +  \frac{g_A^2}{f_\pi^2} \sum_{i<j}  \tilde V_{ij}
\end{eqnarray}
where  $g_A$ is a quark-pion coupling
which scales like $O(N_c^0)$ with the number of colors $N_c$.

\section{$S_N\to S_N^{\rm orb} \times S_N^{sp-fl}$ decomposition of the 
interaction Hamiltonian}
\label{Sec:4}
 
The Hamiltonians $H$ in Eqs.~(\ref{Hamilton}), (\ref{HamiltonGR}) can be 
decomposed into a sum of terms transforming according to irreps of the permutation 
group acting on the spin-flavor degrees of freedom $S_N^{\rm sp-fl}$. 
The operators in Eqs.~(\ref{Hamilton}), (\ref{HamiltonGR}) are 
two-body interactions, of the generic form
\begin{eqnarray}
V = \sum_{1\leq i<j\leq N} {\cal R}_{ij} {\cal O}_{ij}
\end{eqnarray}
where ${\cal R}_{ij}$ acts only on the orbital coordinates of the quarks 
$i,j$, and
${\cal O}_{ij}$ acts only on their spin-flavor degrees of freedom. For 
example, the
spin-spin interaction $V_{ss}$ has ${\cal R}_{ij} = f_0(r_{ij})$ and 
${\cal O}_{ij} = \vec s_i \cdot \vec s_j$. $V$ must be symmetric under
any permutation of the $N$ quarks, but the transformation of
the spin-flavor and orbital factors separately can be more complicated.
We distinguish two possibilities for the transformation of these 
operators under a transposition of the 
quarks $i,j$, corresponding to the two irreps of $S_2$:
i) symmetric two-body operators $V_{\rm symm}$:
$P_{ij} {\cal R}_{ij} P^{-1}_{ij} = {\cal R}_{ij}$ and
$P_{ij} {\cal O}_{ij} P^{-1}_{ij} = {\cal O}_{ij}$.
ii) antisymmetric two-body operators $V_{\rm anti}$: 
$P_{ij} {\cal R}_{ij}P^{-1}_{ij} = -{\cal R}_{ij}$ and
$P_{ij} {\cal O}_{ij}P^{-1}_{ij} = -{\cal O}_{ij}$.
The spin-spin and quadrupole 
interactions $V_{ss}, V_q$ are symmetric two-body
operators, while the spin-orbit interaction $V_{so}$ contains both symmetric and 
antisymmetric components.

In general, the $k-$body operators can be classified into irreps of the
permutation group of $k$ objects $S_k$. For example, there are three classes
of 3-body operators, corresponding to the $S,MS$ and $A$ irreps of $S_3$.

We start by considering the symmetric two-body operators.
The set of all spin-flavor operators ${\cal O}_{ij}$ (and analogous for 
the orbital operators ${\cal R}_{ij}$) 
with $1 \leq i < j \leq N$ 
form a $\frac12 N(N-1)$ dimensional
reducible representation of the $S_N$ group, which contains the following irreps
\begin{eqnarray}
\{ {\cal O}_{ij} \} = S \oplus MS \oplus E
\end{eqnarray}  
We will use as a basis for the operators on the right-hand side the Young 
operator basis supplemented by the phase convention Eq.~(\ref{phitransp}). 
For simplicity we neglect the $E$ irrep in the following explicit example.

The $S$ and $MS$ projections are 
\begin{eqnarray}
{\cal O}^S = \sum_{i<j} {\cal O}_{ij} \ , 
\end{eqnarray}

\begin{eqnarray}
{\cal O}^{MS}_k = \sum_{2\leq j \neq k\leq N} ( {\cal O}_{1j} - {\cal O}_{kj} )\,,
\qquad k = 2, 3, \cdots , N \ , 
\end{eqnarray}

The interaction $V_{\rm symm}$ is symmetric under $S_N$, and its 
decomposition under $S_N \subset S_N^{orb} \times S_N^{sp-fl}$ has the form
{\small
\begin{eqnarray}\label{Oeven}
V_{\rm symm} &=& \frac{2  {\cal R}^S {\cal O}^S }{N(N-1)}+ 
\frac{N-1}{N(N-2)} \sum_{j,k=2}^N  {\cal R}^{MS}_j {\cal O}^{MS}_k M_{jk} \ .
\end{eqnarray}
}
where the matrix $M_{jk}$ is given in Eq.~(\ref{Mmatrix}) in explicit form.

The antisymmetric two-body operators ${\cal O}_{ij}$
form a reducible representation of $S_N$ of dimension$\frac12 N(N-1)$,
which is decomposed into irreps as
\begin{eqnarray}
\{ {\cal O}_{ij} \} = MS \oplus A'
\end{eqnarray}  
and will not be considered further here (for a complete discussion see Ref.~\cite{Pirjol:2007ed}). 

\section{Mass operator - OGE interaction}
\label{Sec:5}

For simplicity we keep only the $S$ and $MS$ operators, and compute their
matrix elements on the $|B\rangle $ states constructed in Eq.~(\ref{MS}). 
The matrix elements of the spin-flavor operators on the basis functions $\phi_i$
are given by
\begin{eqnarray}
\langle \phi_j | {\cal O}^S |\phi_k \rangle &=& 
\langle {\cal O}^S \rangle S_{jk} \\
\label{mswe}
\langle \phi_j | {\cal O}^{MS}_i |\phi_k \rangle &=& 
\langle {\cal O}^{MS} \rangle (1 - \delta_{ji}\delta_{ik}) \ ,
\end{eqnarray}
where $S_{jk}$ was defined in Eq.~(\ref{phinorm}), and 
$\langle {\cal O}^S\rangle, \langle {\cal O}^{MS}\rangle$ are reduced matrix 
elements.
The proportionality of the matrix elements to just one reduced
matrix element follows from the Wigner-Eckart theorem for the
$S_N$ group. The form of the Clebsch-Gordan coefficients is specific
to the $MS$ basis used in Eqs.(\ref{phidef},\ref{chidef}), and can be derived by repeated 
application of Eqs.~(\ref{phitransp}-\ref{phitransp2}) to the states and operators.

Similar expressions apply for the orbital operators ${\cal R}$, up to an
additional complexity 
introduced by the presence of the magnetic quantum numbers of the 
initial and final state orbital basis functions $\chi_p^m$. The 
dependence on $m,m'$ is related to the Lorentz structure of the orbital
operator and can be parametrized by the matrix element of a tensor 
operator which in the present case can be constructed in terms of the 
angular momentum operator $\vec L$.

Inserting these expressions into Eq.~(\ref{Oeven}) and combining all factors 
one finds the general expression for a symmetric 2-body operator
\begin{eqnarray}\label{symm2body}
\frac{\langle B | V_{\rm symm} | B \rangle}{\langle B| B \rangle} = 
\frac{2 \langle {\cal O}^S \rangle
\langle {\cal R}^S \rangle}{N(N-1)} + \frac{\langle {\cal O}^{MS} \rangle
\langle {\cal R}^{MS} \rangle}{N}
\end{eqnarray}
where the contributions of operators transforming in the 
$E$ irrep are neglected.

The reduced matrix elements depend on the precise form of the
interaction. We consider for definiteness the spin-spin interaction $V_{ss}$
 in some detail. For this case the reduced matrix elements of the symmetric operators
are given by
\begin{eqnarray}
\langle {\cal R}^S \rangle &=& 
\frac12 (N-1)(N-2) {\cal I}_{s} + (N-1) {\cal I}_{\rm dir} - {\cal I}_{\rm exc}\\
\langle {\cal O}^S \rangle &=& \langle \Phi(SI) |\frac12 \vec S\,^2 -
\frac38 N |\Phi(SI)\rangle \ . 
\end{eqnarray}
For the ease of comparison with the literature on the $1/N_c$ expansion for 
excited baryons, we expressed the reduced matrix element of the spin-flavor
operator as a matrix element on  the state $|\Phi(SI)\rangle$ where the
excited quark is quark no. 1. 
The reduced matrix elements of the
orbital operator are expressed in terms of the 3 overlap integrals over the
one-body wave functions
\begin{eqnarray}\label{overlap}
{\cal I}_s &=& \int d\vec r_1 d\vec r_2 f_0(r_{12}) |\varphi_s(\vec r_1)|^2 
|\varphi_s(\vec r_2)|^2\\
{\cal I}_{\rm dir} &=& \int d\vec r_1 d\vec r_2 f_0(r_{12}) 
|\varphi_s(\vec r_1)|^2 |\varphi_p^m(\vec r_2)|^2 
\nonumber \\
{\cal I}_{\rm exc} &=& \int d\vec r_1 d\vec r_2 f_0(r_{12}) 
\varphi_s(\vec r_2) \varphi_p^m(\vec r_1) 
\varphi_s^* (\vec r_1) \varphi_p^{*m}(\vec r_2) \nonumber
\end{eqnarray}

The spin-flavor operator transforming in the $MS$ irrep is ${\cal O}_k^{MS} =
(\vec s_1 - \vec s_k) \cdot \vec S$, and the corresponding orbital operator
is ${\cal R}_k^{MS} = \sum_{j=2, j\neq k}^N [f_0(r_{1j}) - f_0(r_{kj})]$.
Their reduced matrix elements are
\begin{eqnarray}\label{OMS}
\langle {\cal O}^{MS} \rangle &=& \frac{1}{N-2}
\langle \Phi(SI) |- \vec S\,^2
+ N \vec s_1 \cdot \vec S_c + \frac34 N |\Phi(SI)\rangle  \nonumber \\
\langle {\cal R}^{MS} \rangle &=& (N-2) 
({\cal I}_{\rm dir} - {\cal I}_{s}) - 2 {\cal I}_{\rm exc} \ . 
\end{eqnarray}
 We
denoted $\vec S_c$ the `core' spin, defined as $\vec S_c = \vec S - \vec s_1$. 
The reduced matrix element of the orbital operator $\langle {\cal R}^{MS} \rangle$
was computed by taking representative values of $i,j,k$ in Eq.~(\ref{mswe}) and
evaluating the integrals.

\subsection{The reduced matrix element $\langle O_{MS}\rangle$ }

We present here the details of the derivation of
the reduced matrix element of a $MS$ spin-flavor operator as a 
matrix element on the CCGL state $\Phi(SI)$ with fixed identity of the `excited' quark
(such as e.g. Eq.~(\ref{OMS})).
As explained in Sec.~\ref{Sec:MSstates}, this state is given by 
\begin{eqnarray}
\Phi(SI) = \frac{1}{\sqrt{N(N-1)}}\sum_{k=2}^N \phi_k\,.
\end{eqnarray}

The matrix element of an $MS$ operator on our $MS$ basis states is given by the Wigner-Eckart
theorem for $S_N$, Eq.(\ref{mswe}).
Summing over the index of the operator, the matrix element on the $\Phi(SI)$ state is
\begin{eqnarray}\label{B3}
\langle \Phi(SI) |\sum_{k=2}^N {\cal O}^{MS}_k |\Phi(SI) \rangle &=& 
(N-2) \langle {\cal O}^{MS} \rangle
\end{eqnarray}
which can be used to express $ \langle {\cal O}^{MS} \rangle$ as a matrix element on the
CCGL-type spin-flavor state $\Phi(SI)$.

The advantage of taking the sum $\sum_{k=2}^N {\cal O}_k^{MS}$ is that it singles
out the quark no.~1, just as in the state $\Phi(SI)$. An explicit calculation gives
\begin{eqnarray}
\sum_{k=2}^N  {\cal O}_k^{MS} =
- 2 \sum_{i<j=1}^N {\cal O}_{ij} + N \sum_{i=2}^N {\cal O}_{1i} 
\end{eqnarray}
for symmetric 2-body operators.
Using this relation, one finds for example for the spin-spin interaction ${\cal O}_{ij}=
\vec s_i \cdot \vec s_j$
\begin{eqnarray}
\sum_{k=2}^N {\cal O}_k^{MS} = -\vec S\,^2 + N \vec s_1 \cdot \vec S_c + \frac34 N
\end{eqnarray}
which gives Eq.~(\ref{OMS}) after combining it with Eq.~(\ref{B3}).

\section{Matching onto the $1/N_c$ expansion}

The spin-flavor structure of the one-gluon exchange interaction 
matches a subset of the operators appearing in the $1/N_c$ expansion of the mass operator 
for orbitally excited states.  Working to order $1/N_c$, the
most general set of operators contributing to the mass of these states is
\begin{eqnarray}
\hat M = c_1 N_c {\bf 1} + c_2 L^i s^i + 
c_3 \frac{3}{N_c} L_2^{ij} g^{ia} G_c^{ja}
+ \sum_{i=4}^8 c_i {\cal O}_i
\end{eqnarray}
The terms proportional to $c_{2,3}$ contribute at order $O(N_c^0)$, and the remaining
operators proportional to $c_{4-8}$ are of order $1/N_c$. A
complete basis of subleading operators
can be chosen as in Ref.~\cite{Carlson:1998vx} and is shown in Table~\ref{tab:2}. The corresponding coefficients that we find matching the quark-quark interactions are given 
in Table~\ref{tab:3} and can be expressed in terms of overlap 
integrals, where ${\cal J , K}$ are similar to the ones given in Eqs.(\ref{overlap}) and originate from the spin-orbit and tensor interaction respectively.  

\begin{table}
\caption{Subleading operators}
\label{tab:2}  
\begin{tabular}{ll}
\hline\noalign{\smallskip}
${\cal O}_4   $ & $ L^i s^i + \frac{4}{N_c+1} L^i t^a G_c^{ia} $ \\
${\cal O}_5   $ & $\frac{1}{N_c} L^i S_c^{i} $ \\
${\cal O}_6   $ & $\frac{1}{N_c} S_c^2$        \\
${\cal O}_7   $ & $\frac{1}{N_c} s^i S_c^i $  \\
${\cal O}_8   $ & $\frac{1}{N_c} L_2^{ij} \{s^i\,, S_c^j\}$    \\
${\cal O}_9   $ & $\frac{1}{N_c} L^i g^{ia} T_c^a $      \\
${\cal O}_{10}$ & $\frac{1}{N_c} t^a T_c^a $   \\
${\cal O}_{11}$ & $\frac{1}{N_c^2} L_2^{ij} t^a \{S_c^i\,, G_c^{ja}\}$     \\
\noalign{\smallskip}\hline
\end{tabular}
\end{table}

\begin{table}
\caption{Matching coefficients}
\label{tab:3}  
\begin{tabular}{lll}
\hline\noalign{\smallskip}
Coeff. & OGE & OBE \\
\noalign{\smallskip}\hline\noalign{\smallskip}
$c_2$ &
$-\frac{g^2 N_c}{4} \Big( 3 {\cal J}_{\rm dir}^s - {\cal J}_{\rm dir}^a \Big)$ &
$-  \frac18 \tilde g_A^2 ( 3 {\cal J}_{\rm dir}^s + {\cal J}_{\rm dir}^a )$
\\
$c_3$ &
$0$ &
$\frac23 \tilde g_A^2 {\cal K}_{\rm dir}$  \\
$c_4$ &
$0$ &
$\frac18 \tilde g_A^2 ( 3 {\cal J}_{\rm dir}^s + {\cal J}_{\rm dir}^a )$  \\
$c_5$ &  
$- \frac{g^2 N_c}{4} \Big( 3 {\cal J}_{\rm dir}^s  + {\cal J}_{\rm dir}^a \Big)$
& $0$ \\
$c_6$ &  $-\frac{g^2 N_c}{4} {\cal I}_s$ & $-\frac14 \tilde g_A^2  {\cal I}_{\rm s}$ \\
$c_7$ & $-\frac{g^2 N_c}{2} {\cal I}_{\rm dir}$        &
$-\frac14 \tilde g_A^2 {\cal I}_{\rm dir}$ \\
$c_8$ & $-\frac{g^2 N_c}{2}  {\cal K}_{\rm dir}$        & $0$ \\
$c_9$ &     $0$    & $\frac12 \tilde g_A^2 ( 3 {\cal J}_{\rm dir}^s - {\cal J}_{\rm dir}^a )$ \\
$c_{10}$ &      $0$   & $-\frac14 \tilde g_A^2 {\cal I}_{\rm dir}$\\
$c_{11}$ &   $0$      & $0$ \\
\noalign{\smallskip}\hline
\end{tabular}
\end{table}

The complete explicit calculation \cite{Pirjol:2007ed} confirms the $N_c$ power counting rules of Ref.~\cite{Goity:1996hk,Carlson:1998vx}, in particular the 
leading order $O(N_c^0)$ contribution to the mass coming from the 
spin-orbit interaction given by $c_2$
confirms in a direct way the prediction 
of the breaking of the $SU(4)$ spin-flavor symmetry at leading order in $N_c$
\cite{Goity:1996hk}.
The nonrelativistic quark model with gluon mediated quark interactions displays
the same breaking phenomenon.

 The two-body quark interactions considered
here produce one-, two- and three-body (${\cal O}_{17} = \frac{1}{N_c^2}L_2^{ij}
\{ S_c^i\,, S_c^j \}$ of Ref.~\cite{Carlson:1998vx}, which correctly 
appears at order $O(1/N_c^2)$ ) effective 
operators in the $1/N_c$ expansion.

Another important conclusion following from this calculation is that 
operators with nontrivial permutation symmetry are indeed required by a correct 
implementation of the $1/N_c$ expansion.

A distinctive prediction of the one-gluon exchange potential is the vanishing
of the coefficient $c_3$ of one of the leading order operators. On the other 
hand, in the one-boson exchange model,
the coefficients of the two leading operators can be of 
natural size.
There are predictions for the vanishing of the coefficients
of some subleading operators, and also one
relation between the coefficients of the leading and subleading operators
$c_2= -c_4 $. The Goldstone boson interaction also generates the
three-body operator, ${\cal O}_{17} $ of
Ref.~\cite{Carlson:1998vx}, which correctly appears at 
order $O(1/N_c^2)$.

The coefficients of the leading order operators $c_{1,2,3}$ have been
determined in  \cite{Pirjol:2003ye} from a fit to the masses of the nonstrange 
$L=1$ baryons, working at leading order in $1/N_c$. 
The results depend on the assignment of the observed baryons into the irreps
of the contracted symmetry (towers). There are four possible assignments, but
only two of them are favored by data. These two assignments give the
coefficients
\begin{eqnarray}
&& \!\!\!\!\!\!\!\!\!\!\!\!\!\! \mbox{assignment 1: } \nonumber \\
&& \!\!\!\!\!\!\!\!\!\!\!\!\!\! c_2^{(0)} = 83\pm 14 \mbox{ MeV}\,,\qquad c_3^{(0)} = -188\pm 28 \mbox{ MeV} \\
&& \!\!\!\!\!\!\!\!\!\!\!\!\!\! \mbox{assignment 3: } \nonumber \\  
&& \!\!\!\!\!\!\!\!\!\!\!\!\!\! c_2^{(0)} = -12\pm 16 \mbox{ MeV}\,,\qquad c_3^{(0)} = 142\pm 38 \mbox{ MeV} 
\end{eqnarray}
Comparing with the predictions from the gluon mediated
quark-quark interactions, we see that there is no evidence in the data for
a suppression of the coefficient $c_3$ relative to $c_2$. For both assignments,
the coefficient $c_3$ is sizeable, a situation which favors the
Goldstone boson exchange model, or at least indicates that 
some kind of flavor dependent effective interactions cannot be neglected.

\section{Conclusions}

We presented a general procedure to match quark models to 
the $1/N_c$ expansion of QCD, using as an example the mass 
operator for orbitally excited baryons. The transformation properties of the states and operators under the permutation
group $S_N$ allow  to parameterize our ignorance of the spatial dependence 
of the wave functions in terms of only a few reduced matrix elements.

This approach is similar to the method used for
$N_c=3$ in Ref.~\cite{Collins:1998ny} to compare the predictions
of  nonrelativistic quark models in a form independent of the orbital 
wave functions.
In addition to extending this result to arbitrary $N_c$, we also extract 
here the spin-flavor structure in an operator form, which allows to make an 
explicit comparison of two different models before taking the full matrix elements.

The results obtained for the mass operator match precisely the operators appearing
in the $1/N_c$ expansion \cite{Goity:1996hk,Carlson:1998vx},
up to new contributions, not considered previously, transforming in the $E$ 
and $A'$ irreps of $S_N^{\rm sp-fl}$. We confirmed by an explicit calculation the $N_c$ power counting rules of
Refs.~\cite{Goity:1996hk,Carlson:1998vx}. 
In particular we proved that the effective spin-orbit 
interaction is of leading order
$O(N_c^0)$. 

The transformation properties under $S_N$ also imply that the  decomposition into
core and excited quark operators used previously in the literature on the
subject is necessary to obtain the most general expression for the 
mass operator.

The two models considered here induce different hierarchies among the effective 
constants of the general $1/N_c$ expansion.
Comparing with existing fits to the masses of nonstrange $L=1$ excited baryons,
we find that flavor dependent interactions cannot be neglected, and
may be necessary to supplement the gluon exchange model. This is 
in line with the chiral quark picture proposed in 
Ref.~\cite{Manohar:1983md}.

%
%

\end{document}